\documentclass{article}
\usepackage{graphicx}
\usepackage{amsmath,amssymb,yhmath,bm}
\title{ Minisuperspace quantum supersymmetric cosmology\\ (and its hidden hyperbolic Kac-Moody structures).}
\author{Thibault Damour\\Institut des Hautes \'Etudes Scientifiques\\
35, route de Chartres\\
F 91440 Bures sur Yvette, France\\
E-mail: damour@ihes.fr\and
Philippe Spindel\\M\'ecanique et Gravitation, UMONS\\
20, place du Parc\\
7000 Mons, Belgique\\
E-mail: spindel@umons.ac.be}

\begin{document}

\maketitle
\begin{abstract}
This work summarises recent progress \cite{Damour:2013eua, Damour:2014cba} obtained by the  mini-superpspace quantization of $\mathcal N=1$, $d=4$ supergravity, formulated in the framework of the Bianchi IX cosmological model. The emphasis is put on three main results : the completeness of the solution space obtained, the elements suggesting a hidden Kac-Moody structure of the theory and those leading to conjecture an avoidance of the cosmological singularity by some branches of the wave function of the Universe.
\end{abstract}
{\par\noindent}{\bf keywords : }Quantum cosmology, supergravity,  quantum gravity, Kac-Moody algebras


\section{Some Motivations}

In the framework of Einstein gravity, under very general assumptions, several theorems \cite{HE} predict ineluctable occurrence of cosmological singularities. These singularities
 reflect the inadequacy of a classical description of the physics of the first instants of the Universe. The relevant theory --- maybe a workable string field theory --- is not yet known. However, we may think that the era since which  physics is reasonably well described by a metric governed by  general relativity is preceded by an epoch where a quantized version of it is relevant. Of course this is only a {\it pis aller}, comparable to the quantization of the hydrogen atom or to the Fermi model that can be seen as approximations to quantum electrodynamics or quantum non-abelian gauge theories.
{\par\noindent}
On the other hand, a study  of classical cosmological solutions in the vicinity of their singularities \cite{BKL,Misner:1969hg} has shown that in a generic case the dynamics of the system becomes chaotic. Even homogeneous cosmologies like the Bianchi VIII or IX models approximately evolve as a succession of periods described by Kasner like metrics : 
$$ds^2=-dt^2+\sum_kt^{2\,p_k}(\theta^k)^2$$
separated by collisions.
Here $\{\theta^k\}$ is a set of $SO(2,1)$ or $SU(2)$ invariant forms and  $\{p_k\}$ the set of so-called Kasner exponents. These exponents are constants during each period, but randomly  reshuffled by the collision processes. The evolution of the classical geometry of a general non-homogeneous model appears as an ensemble of independent Bianchi VIII or IX models attached to each point of a 3-dimensional section of the spacetime.
{\par\noindent}
A finer analysis of this dynamics has put into evidence some surprising elements, suggesting a possible  hidden symmetry of gravity or, if not, at least of the solution space of the equations of the theory. Indeed, for all models of supergravity it was  shown that there exists  a correspondence between the dynamics of some of their degrees of freedom and the dynamics of a spinning particle moving on a (formal) coset manifold built from the quotient of a Kac-Moody group by its maximally compact subgroup\cite{Damour:2001sa,Damour:2000hv,Damour:2002et}. For instance, the well known
BKL oscillatory behaviour  of the diagonal components of a generic, inhomogeneous
Einsteinian vacuum metric in $d = 4$ was found to be equivalent to a billiard motion within the Weyl
chamber of the rank 3 hyperbolic KacÐMoody algebra $AE_3$. Similarly, the generic BKL like
dynamics of the bosonic sector of maximal supergravity (considered either in $d=11$, or,
after dimensional reduction, in $4\leq d \leq 10$) leads to a chaotic billiard 
motion within the Weyl
chamber of the rank-10 hyperbolic KacÐMoody algebra $E_{10}$. For reviews on this, see for instance  [\cite{Damour:2002et}, \cite{Henneaux:2007ej}].
{\par\noindent}
Hereafter we shall illustrate these considerations in the framework of $\mathcal N=1$, $d=4$ simple supergravity. Our strategy is a generalisation of the one used in Ref.~[\cite{Damour:2011yk}].
After having dimensionally reduced this theory on a $SU(2)\equiv S^3$ manifold (the arena of the Bianchi IX cosmological model), we quantize it {\it \`a la Dirac}\/. We have so obtained an exhaustive description of the solution space of the corresponding Wheeler DeWitt constraint equations. In particular, new solutions, overlooked in all previous works we know on the subject, have been encountered. We also have shown the existence of elements characteristic of an underlying Kac-Moody algebra hidden in the algebra of the quantum operators that are at the basis of all the quantum dynamics. These operators generate a spinorial extension\cite{Damour:2005zs,deBuyl:2005mt,Damour:2009zc} of the $AE_3$ Weyl group. In the limit of small wavelengths they lead to a chaotic quantum evolution of the Universe near the singularity. They also provide arguments  allowing us to conjecture that some components of the wave function of the Universe avoid the cosmological singularity.
\section{The Minisuperspace Model}
Specifically the model we consider rests on the supergravity Lagrangian density :
\begin{equation}\label{Ltot}
 {\mathcal L}\strut_{Tot}=\theta\left[L_{E-H}+\ring L_{3/2}+\frac 18 {T^{\hat \alpha}} T_{\hat \alpha}-\frac1{16}T^{{\hat \alpha}\hat \beta\hat \gamma}T_{\hat \gamma\hat \beta{\hat \alpha}}-\frac 1{32}T^{{\hat \alpha}\hat \beta\hat \gamma}T_{{\hat \alpha}\hat \beta\hat \gamma} \right]
\end{equation}
where
\begin{equation}
 T_{{\hat \alpha}\hat \beta\hat \gamma}:=\overline\psi_{\hat \beta}\gamma_{\hat \alpha}\psi_{\hat \gamma}\qquad,\qquad T_{\hat \alpha}=\overline\psi_{\hat \alpha}\gamma^{\hat \beta}\psi_{\hat \beta}\qquad .
\end{equation}
expressed in terms of a co-frame $\{\theta ^{\hat \mu}\vert\, \mu=0,\dots,3\}$ and a Majorana vector-spinor (spin $3/2$) field $\psi_{\hat \alpha}$.
{\par\noindent}
Here $L_{E-H}=1/2\,\ring R$ and $\ring L_{3/2}=-1/2 \epsilon^{\hat \alpha\hat\beta\hat\gamma\hat\delta}\overline\psi_{\hat\alpha}\gamma_5\gamma_{\hat\beta}\nabla_{\hat\gamma}\psi_{\hat\delta}$ are respectively the Einstein-Hilbert and the Rarita-Schwinger Lagrangians built with the metric 
 \begin{equation}
ds^2=\eta_{\hat \alpha\hat\beta}\theta^{\hat \alpha}\theta^{\hat\beta}\qquad .
\end{equation}
and the associated Levi-Civita connection.  An important difference between our approach and numerous previous ones (see for instance Refs~[\cite{death:1993ki} -- \cite{vargas-moniz:2010fu}]) is that we fix from the beginning the Lorentz gauge by univocally specifying a vierbein  expressed as (time dependent) combinations of the 1-form $dt$ and the three invariant forms on $SU(2)$ : $\tau^a (x)$   (which satisfy $d\tau^a+\epsilon^a_{\phantom{a}b\,c}\tau^b\wedge\tau^c=0$). In terms of them the metric reads :
\begin{equation}
g_{\mu\,\nu}dx^\mu dx^\nu=-N^2(t)\,dt^2+g_{a\,b}(\tau^a(x)+N^a(t)\,dt)(\tau^b(x)+N^b(t)\,dt)\qquad .
\end{equation}
We parametrize {\it \`a la Gauss} the Euclidean metric matrix components 
\begin{equation}
g_{bc}(t)=\sum_{\hat a}e^{-2\beta^{\hat a}(t)}S^{\hat a}_{\phantom{a}b}[\varphi^m(t)]S^{\hat a}_{\phantom{a}c}[\varphi^n(t)]
\end{equation}
 in terms of its (time dependent) eigenvalues $e^{-2\beta^{\hat a}}$ and the orthogonal matrix $S^{\hat a}_{\phantom{a}b}[\varphi^c(t)]$ (depending on three time dependent Euler
 angles denoted $\{\varphi^a(t)\}$)  that diagonalises it. This decomposition allows one to fix the coframe $\theta^{\hat\alpha} = h_{ \ \ \mu}^{\hat\alpha} \, dx^{\mu}$ as :
\begin{equation}
\theta^{\hat 0} = N(t) dt\qquad,\qquad \theta^{\hat a} = {\sum_b} \, e^{-\beta^{\hat a} (t)} S_{ \ \ b}^{\hat a} (\varphi^c(t))(\tau^b (x) + N^b(t) dt)\label{BIXforms}\qquad .
\end{equation}
Accordingly we do not have to worry about the six local Lorentz constraints. This simplification is the main technical ingredient that distinguishes our work  from all the previous ones on the subject, and allows us to  completely solve  the quantization of the reduced problem. To go ahead we also introduce the frame angular velocity $(\dot S\,S^{-1})^{\hat a}_{\phantom{a}\hat b}=\epsilon^{\hat a}_{\phantom{ a}\hat b\,\hat c}\omega^{\hat c}$. Moreover we redefine the sixteen fermionic variables as $\Psi_{\hat \alpha}:=g^{1/4}  \psi_{\hat\alpha}$ where $g:=(\det g_{a\,b})=\exp [-1/2(\beta^1+\beta^2+\beta^3)]$ and finally put $\Psi'_{\hat 0}=\Psi_{\hat 0}-\gamma_{\hat 0}\,\gamma^{\hat a}\Psi_{\hat a}$.  In terms of these sixteen fermionic variables and ten bosonic ones ($N,\ N^a,\ \beta^a\ \phi^a$) we may rewrite the Lagrangian density into a Hamiltonian form, a first step for quantization (from now we use units such that $\hbar=1$, $c=1$, $G=2\,\pi$ (so that $8\,\pi\,G=16\,\pi^2=\mbox{vol.}(S^3)$) 
:
\begin {equation}\label{LH}
\mathcal L_H=\pi_a\,\dot\beta^a+p_{\omega^a}\,\omega^a+\frac 12 \overline\Psi_{\hat a}\gamma^{\hat a\hat 0\hat b}\dot \Psi_{\hat b}+\overline\Psi'_{\hat 0}\,{\mathcal S}-N\,\sqrt{g}\,H-N^a\,H_a
\end{equation}
We observe the occurrence of eight Lagrange multipliers. The four components of the spinor $\Psi'_{\hat 0}$, $N\,\sqrt{g}$ and $N^a$, respectively multiplying the four generators of supersymmetry ${\mathcal S}$ (here we didn't make the spinor indices explicit), and the four spacetime reparametrizations generators $H,\, H_a$ (whose explicit expression are provided  in Ref. [\cite{Damour:2014cba}]). As usual the kinetic term of the Lagrangian fixes the (anti)commutation rules of the (fermionic) bosonic variables. For the latter we adopt a Schr\"odinger representation on a space of functions depending on the $\beta^a$ and $\varphi^a$ variables; the conjugate momentum operators being represented by  : $\hat \pi_a=-i\partial_{\beta^a}$, $\hat p_{\varphi ^a}=-i\partial_{\varphi^a}$ (the operators $\hat p_{\omega^a}$ being $\varphi^a$-dependent linear combinations of the latter ones; see Ref. [\cite{Damour:2011yk}] for a discussion od this point). 
For the fermionic variables, we reads from their kinetic terms in Eq. [\ref{LH}] that the corresponding operators have to obey the anticommutation relations :
\begin{equation}\label{ACrelPsi}
\{\widehat \Psi_{\hat a}^A,\,\widehat\Psi_{\hat b}^B\}=-\frac i2\left(\gamma_{\hat b}\gamma_{\hat a}\gamma^{\hat 0}\mathcal C^{-1}\right)^{A\,B}
\end{equation}
where $\mathcal C$ is a charge conjugation matrix : $\mathcal C\gamma_\mu=(\mathcal C\gamma_\mu)^T$.
\section{Classical Evidences for a Hidden Kac-Moody Structure}
When reduced on $SU(2)$, using the coframe defined bys Eqs [\ref{BIXforms}], the purely gravitational part of the Lagrangian [\ref{Ltot}] reduces to :
 \begin{eqnarray}
8\pi G \, \ring{\mathcal L}_{\rm EH}&=&\frac 12 N\, \sqrt{g}\,\ring R\nonumber  \\
&=&\frac 1 {N } \, e^{-\sum_a\beta^a}\left\{-(\dot \beta^1\dot \beta^2+\dot \beta^2\dot \beta^3+\dot \beta^3\dot \beta^1)+(N^{\overline 1}+\omega^1)^2\sinh^2[\beta^2-\beta^3]\right.\nonumber \nonumber  \\
&&\left.+(N^{\overline 2}+\omega^2)^2\sinh^2[\beta^3-\beta^1]+(N^{\overline 3}+\omega^3)^2\sinh^2[\beta^1-\beta^2]\right\}\nonumber \nonumber  \\
&&-{N \,e^{-\Sigma_a\beta^a}}\,\left\{ \frac14 \, \sum_b e^{-2(2\beta^b-\Sigma_a\beta^a)}-\frac 12 \sum_b e^{2\beta^b}\right\} \qquad .\nonumber
\end{eqnarray}
This result displays several structures. Firstly, from the kinetic part of the scale factor exponents  we recognise the reduced form of the DeWitt metric~:
\begin{equation}
\left(G_{a\,b}\right)=
\left(
\begin{array}{ccc}
 0 &-1   & -1  \\
  -1&  0 &   -1\\
 -1 &-1   &  0 
\end{array}
\right)\qquad ,
\end{equation}
a 3-dimensional Lorentzian metric, with signature $(-,+,+)$. We also notice that the rotational kinetic term presents singularities on the symmetry walls :    
$
w^s_1(\beta):=\beta^2-\beta^1=0\ ,\, w^s_2(\beta):=\beta^3-\beta^2=0\ ,\,\ w^s_3(\beta):=\,\beta^3-\beta^1=0\ .$.
Finally, the last, potential term exhibits the following ``gravitational walls" : $w^g_1(\beta):=2\,\beta^1 =0$, $w^g_2(\beta):=2\,\beta^2 =0$, $w^g_3(\beta):=2\,\beta^3 =0$.
{\par\noindent}
There is a remarkable connection between this geometrical structure of the $\beta^a$-space and the Kac-Moody algebra $AE_3$. Actually the connection extends to all supergravity theories (see Ref. \cite{Henneaux:2007ej} for a pedagogical review on this), but here we shall restrict ourselves to the specific case we are considering : the simple $\mathcal N=1$, $d=4$ supergravity. The Kac-Moody algebras that are relevant for our purpose are defined from the data of an invertible $n\times n$ Cartan matrix $A$ whose entries satisfy the conditions
\begin{equation}\label{Cartanmat}
\forall\ i,\, j =1,\dots,n \ :\ A_{i\,i}=2\quad,\quad A_{i\,j}\in \mathbb Z_-\ (i\neq j)\quad,\quad A_{i\,j}=0\ \mbox{iff}\ A_{j\,i}=0\quad. 
\end{equation}
Moreover we also require the matrix $A$ to be symmetrizable {\it i.e.} that it exists a diagonal matrix $D=\mbox{diag.}(\epsilon_1,\epsilon_2,\dots,\epsilon_n)$ and a symmetric matrix $S$ such that $A=D\,S$.
The Kac-Moody algebra itself is obtained from the Chevalley-Serre relations, which intertwine $n$ $sl(2)$ algebras as follows. Let us consider a set of $n$ triplet of generators $\{e_i,\ f_i, \, h_i: [h_i,e_i]=2\,e_i,\  [h_i,f_i]=-2\,f_i,\  [e_i,f_i]=h_i,\ i=1,\dots, n\}$  and extend their commutation relations to
\begin{equation}
 [h_i,h_j]=0\ ,\  [h_i,e_j]=A_{ij}\,e_i\ ,\  [h_i,f_j]=-A_{ij}\,f_i \ ,\  [e_i,f_j]=\delta_{ij}\,e_i\ ,
\end{equation}
plus the so-called Serre relations
\begin{equation}
\mbox{ad}_{e_i}^{1-A_{ij}}(e_j)=0=\mbox{ad}_{f_i}^{1-A_{ij}}(f_j)\quad,\quad (i\neq j)\qquad.
\end{equation}
Unless $S$ is a positive-definite matrix, the algebra generated by these relations leads to an infinite dimensional simple algebra. As in the case of finite simple Lie algebras, 
the matrix elements of $S$ lead to a (in general non-euclidean) metric structure on the root space, and on the algebra itself. On the simple roots it reads :
\begin{equation}
\langle\alpha_i\vert\alpha_j\rangle:=S_{i\,j}\qquad .
\end{equation}
By duality it leads to a  scalar product on the Cartan subalgebra, the subspace generated by the $h_i$
\begin{equation}
\langle h_i\vert h_j\rangle=\epsilon_i\,\epsilon_j\,S_{i\,j}\qquad,
\end{equation}
and by requiring invariance $\langle[x,y]\vert z\rangle=\langle x\vert [y,z]\rangle$ it can be extended to the all algebra.
{\par\noindent}
On the other hand, the algebra  is invariant with respect to the ``exchange" of the $e_i$ and $f_i$ generators : the Chevalley involution, given by 
\begin{equation}
\omega(e_i)=-f_i\qquad,\qquad \omega(f_i)=-e_i\qquad,\qquad \omega(h_i)=-h_i\qquad .
\end{equation}
This automorphism of the algebra allows one to define a ``maximally compact" subalgebra, the subalgebra of the elements fixed by the involution, generated by the elements $x_i:=e_i-f_i$. The scalar product $\langle\cdot\vert\cdot\rangle$ is negative definite on this subalgebra.\vspace{1.5mm}\strut
{\par\noindent}
The link between the dynamical system we study and the Kac-Moody algebra $AE_3$ is obtained as follows. On the Lorentzian $\beta$-space, the symmetry and gravitational walls define (hyper-)planes.  Near the cosmological singularity (the asymptotic regime where  $\beta^0:=\sum_a\beta^a(t)$ becomes large) these walls become sharp. The classical behaviour of the geometry is well approximated by the motion of a free spinning particle (whose position is given by $\beta^a(t)$) moving on null lines in the $(2+1)$-dimensional Lorentzian space of metric $G_{ab}$ and bouncing on the symmetry and gravitational walls that bound a region in which the particule motion becomes asymptotically confined.
{\par\noindent}
Remarkable facts are that these structures are also encountered in the root space of the $AE_3$ Kac-Moody algebra. The planes defining the walls are those corresponding to the simple roots of the algebra. The regions between them can be identified with Weyl chambers and the reflection on the walls as the (spinorial) Weyl group reflections\cite{Damour:2009zc}. Moreover a truncated subset of the supergravity equations of motion are the same as those of a $\sigma$-model defined on a coset obtained by quotienting the (formal) Kac-Moody group by its maximally compact subgroup\cite{Damour:2002cu}.
\section{Dirac Quantization }
The quantization of the fermionic field is greatly simplified by the introduction of new gravitino variables\cite{Damour:2009zc} : {\small $\Phi^a_A=\sum_B\gamma^{\hat a}_{A\,B}\Psi^B_{\hat a}$}, in terms of which the action becomes\,:
\begin{equation}
S = \int dt \left[ \pi_a \, \dot\beta^a + p_{\omega^a} \, \omega^a + \frac{i}2 \, G_{ab} \, \Phi_A^a \, \dot\Phi_A^b + \bar\Psi'^A_{{\hat 0}} \, {{\mathcal S}}_{A} - \tilde N H - N^a H_a \right]\qquad ,
\end{equation}
involving, for the fermionic kinetic term, again the $AE_3$ root-space metric. This reflects once more the supersymmetry invariance of this action.
Their corresponding operators have to obey the anticommutation relations :
\begin{equation}\label{Cliff84}
\{\widehat\Phi^a_A,\,\widehat\Phi^b_B\}=G^{a\,b}\,\delta_{A\,B}\qquad ,
\end{equation}
which constitute  the Clifford algebra ${\rm Spin} \, (8^+ , 4^-)$. Accordingly, the $\widehat\Phi^a_A$ operators will be represented by appropriate $64\times 64$ matrices acting on a 64-dimensional spinor of $SO(8,4)$ : the wave function of the Universe $\bm  \Psi=\Psi[\beta,\varphi]$, whose components depend {\it a priori} on the $\beta^a$ and $\varphi^a$.
{\par\noindent}
The  diffeomorphism constraints greatly simplify the functional form of $\bm \Psi$. They reduce to : 
\begin{equation}
\widehat  p_{\varphi^a} \, \Psi = -i \, \frac{\partial}{\partial \varphi^a} \, {\bm \Psi} = 0
\end{equation}
and thus imply that $\bm \Psi$ is $\varphi^a$ independent. The resulting
 ``s-wave" function $\Psi (\beta^a)$ must still be submitted to ``five" constraints : 
\begin{equation}
\widehat {{\mathcal S}}_A (\widehat \pi , \beta , \widehat \Phi) \, {\bm \Psi} = 0 \, , \quad \widehat  H (\widehat \pi , \beta , \widehat \Phi) \, {\bm \Psi} = 0\qquad .
\end{equation}
These constraints constitute a (heavily overdetermined) system of $ 4 \times 64 + 64=320$ partial differential equations for the $64$ components of $\bm \Psi$, 64 functions $\Psi_{\sigma} (\beta^1 , \beta^2 , \beta^3)$.
{\par\noindent}
 Actually there exists a (natural)  ordering of the operators $\widehat{\mathcal S}_A$ and $\widehat H$ such that 
they verify the (anti)commutation relations
\begin{eqnarray}\label{SASB}
&&\widehat {{\mathcal S}}_A \, \widehat {{\mathcal S}}_B + \widehat {{\mathcal S}}_B \, \widehat {{\mathcal S}}_A = 4 \, i \, \underset{C}{\sum} \, \widehat  L_{AB}^C (\beta) \, \widehat {{\mathcal S}}_C + \frac12 \, \widehat  H \delta_{AB}\\
&&[\widehat {{\mathcal S}}_A , \widehat  H] = \widehat  M_A^B(\beta) \, \widehat {{\mathcal S}}_B \end{eqnarray}
the second ones being a trivial consequence of the first. Thus it ``only" suffices to solve a system of 256 partial first order equations. To this aim we introduce a diamond structure on the space of solutions of the constrain equations by defining ``creation" and ``annihilation" operators, obtained from chiral projection of the fermionic operators\cite{Damour:2014cba} ($b^k\propto (1-i\,\gamma^5)_A^{\ B}{\widehat\Phi}^k_B$, $\tilde b^k\propto (1+i\,\gamma^5)_A^{\ B}{\widehat\Phi}^k_B$ ; here we denote the hermitian conjugate of an operator by a tilde) : 
\begin{equation}
b^k_{+}={\widehat\Phi}^k_1+i\,{\widehat\Phi}^k_2\quad,\quad
b^k_{-}={\widehat\Phi}^k_3-i\,{\widehat\Phi}^k_4\quad,\quad
\tilde b^k_{+}={\widehat\Phi}^k_1-i\,{\widehat\Phi}^k_2\quad,\quad
\tilde b^k_{-}={\widehat\Phi}^k_3+i\,{\widehat\Phi}^k_4\quad,
\end{equation}
which, as a consequence of the anticommutation relations [\ref{Cliff84}], have as only non-vanishing anticommutators : \begin{equation}
\{b_\epsilon^k\,,\,\tilde b_{\epsilon'}^l\}=2\,G^{kl}\delta_{\epsilon\,{\epsilon'}}\,\mbox{Id}_{64}\qquad .
\end{equation}
Expressed in terms of these operators, the supersymmetry constraints have the form
\begin{equation}\label{Seps}
\widehat {\mathcal S}_\epsilon = \frac i2\partial_{\beta_k}b^k_\epsilon+\alpha_k\,b^k_\epsilon+\frac 12 \mu_{klm}\,B^{klm}_\epsilon+\rho_{klm}C^{klm}_\epsilon+\frac 12 \nu_{klm}\,D^{klm}_\epsilon
\end{equation}
where
\begin{equation}
B^{klm}_\epsilon= b_\epsilon^k\, b_\epsilon^l\,\tilde b_\epsilon^m-G^{lm}\,b^k_\epsilon+G^{km}b^l_\epsilon,\ 
C^{klm}_\epsilon= b_{\epsilon}^k\, b_{-\epsilon}^l\,\tilde b_{-\epsilon}^m-G^{lm}\,b^k_\epsilon,\ 
D^{klm}_\epsilon= b_{-\epsilon}^k\, b_{-\epsilon}^l\,\tilde b_\epsilon^m
\end{equation}
and all the tensor components $\alpha_k,\ \mu_{klm}\,\ \rho_{klm},\  \nu_{klm}$ are $\beta$-dependent but purely imaginary, as they must be to insure hermiticity of the operators. Their explicit forms are displayed in Ref. [\cite{Damour:2014cba}].
Let us also introduce a fermion number operator $\widehat N_F$ : 
\begin{equation}
\widehat N_F:=G_{ab}\tilde b^a_+b^b_+ + G_{ab}\tilde b^a_-b^b_-=\frac12G_{ab}\overline{\widehat \Phi}\strut^a\gamma^{{\hat 1}{\hat 2}{\hat 3}}{\widehat \Phi}^b+3
\end{equation}
that will be a  key tool to describe the structure of the solution space of the quantum constraint equations.

\section{Solution Space Structure}
The wave function we are looking for has to satisfy the four constraint equations~:
\begin{equation}
\widehat{\mathcal S}_\epsilon\,\bm \Psi=0=\widetilde{\widehat{\mathcal S}}_\epsilon\,\bm \Psi\qquad .
\end{equation}
By noticing that $ \widehat{\mathcal S}_\epsilon\,\widehat N_F=(\widehat N_F-1)\,\widehat{\mathcal S}_\epsilon$ and  $\widetilde{\widehat{\mathcal S}}_\epsilon\,\widehat N_F=(\widehat N_F+1)\,\widetilde{\widehat{\mathcal S}}_\epsilon$ we may easily prove that the solution space splits according to the fermion number eigenvalues $N_F$, that runs from 0 to 6. This leads to a diamond structure of the solution space. 
Starting from a unique lower state built on $\Psi_0$ such that $b^k_\epsilon \Psi_0=0$ we decompose the 64-dimensional subspace into seven levels : 
\begin{itemize}
\item{Level 0} : $\Psi_0$ (dim=1)
\item{Level 1} : $\{\tilde b^k_+,\ \tilde b^l_-\}\Psi_0$  ( dim=6=$2 \times 3$)
\item{Level 2} : $\{\tilde b^{(k}_+\tilde b^{l)}_-,\ \tilde b^{[k}_+\tilde b^{l]}_+,\ \tilde b^{[k}_+\tilde b^{l]}_-,\ \tilde b^{[k}_-\tilde b^{l]}_-\}\Psi_0$ (dim=15= 6+3+3+3)
\item{Level 3} :
$\{\frac 12\epsilon^{[a}_{\phantom{[a}k\,l}\tilde b^{m]}_-\tilde b^k_+\tilde b^l_+,
\ \tilde b^1_-\tilde b^2_-\tilde b^3_-,
\ \frac 12\epsilon^{(a}_{\phantom{[a}k\,l}\tilde b^{m)}_-\tilde b^k_+\tilde b^l_+\}\Psi_0,$ (dim=20=$2\times  (3+1+ 6)$) 
\item{Level 4} : \\$\{\frac 14\epsilon^{(a}_{\phantom{(a}k\,l}\epsilon^{b)}_{\phantom{b)}p\,q}\tilde b^{k}_-\tilde b^l_-\tilde b^{p}_+\tilde b^q_+,\frac 14\epsilon^{[a}_{\phantom{[a}k\,l}\epsilon^{b]}_{\phantom{b]}p\,q}\tilde b^{k}_-\tilde b^l_-\tilde b^{p}_+\tilde b^q_+,\tilde b^{k}_-\tilde b^1_+\tilde b^2_+\tilde b^3_+,\tilde b^{k}_+\tilde b^1_-\tilde b^2_-\tilde b^3_-\}\Psi_0$ (dim=15= 6+3+3+3)
\item{Level 5} : $\{\frac 14\epsilon^{(a}_{\phantom{(a}k\,l}\tilde b^{k}_-\tilde b^l_-\tilde b^{1}_+\tilde b^2_+\tilde b^3_+,\frac 14\epsilon^{(a}_{\phantom{(a}k\,l}\tilde b^{k}_+\tilde b^l_+\tilde b^{1}_-\tilde b^2_-\tilde b^3_-\}\Psi_0$ (dim=6 = 3+3)
\item{Level 6} : $\tilde b^1_+\tilde b^2_+\tilde b^3_+\tilde b^1_-\tilde b^2_-\tilde b^3_-\Psi_0$ (dim=1)
\end{itemize}
More explicitly, denoting by $x:=e^{2\beta_1},\ y:=e^{2\beta_2},\ z:=e^{2\beta_3},$ and writing the \underline{level zero} solution as $\bm \Psi_{(0)}=f\,\Psi_0$, we 
find a {\it unique}  (modulo rescaling)  solution $f$ from a set of three (compatible) equations :
\begin{equation}
\frac i2\partial_{\beta_k}f-\phi_k f=0\  \mbox{where}\  
\phi_k=-i\,\left\{\frac 12-\frac 1{2\,x}-\frac 3 8\frac{y(x-z)+z(x-y)}{(x-y)(x-z)}, cyclic\ perm.\right\}\  .
\end{equation}
Its solution, depending on one normalisation constant, is :
\begin{equation}\label{f0}
f=f_0 \,Exp[-\frac 12(\frac 1x+\frac 1y+\frac 1z)](x\,y\,z)^{-\frac 54} \left((x-y)(y-z)(z-x)\right)^{3/8}\quad .
\end{equation}
This result is close to previously obtained solutions, but differs by  factors vanishing on the symmetry walls.\vspace{1.5mm}
{\par\noindent}
At \underline{level 1}, we write the wave function as 
$\bm \Psi_{(1)}=\sum_{k,\,\epsilon} f_k^\epsilon\,\tilde b^k_\epsilon\,\Psi_0$. It a priori depends on six unknown functions. They are determined by the equations :
\begin{equation}
{\mathcal S}_\epsilon\bm \Psi_{(1)}=0\Leftrightarrow
\frac i2 \partial_{\beta_k}\, f^k_\epsilon+\varphi_k\,f^k_\epsilon=0\qquad ,
\end{equation}
and 
\begin{equation}
\tilde{\mathcal S}_\epsilon \bm \Psi_{(1)}=0\Leftrightarrow\left\{\begin{array}{ll}
\tilde\nu_{[kl]}^{\phantom{[kl]}m}\,f_m^\epsilon=0\vspace{1.5mm}&\quad,\\
\left(\frac i2 \partial_{\beta_{[k}}\, f_{l]}^\epsilon+\tilde\varphi_{[k}\,f_{l]}^\epsilon-2\,\tilde \mu_{[kl]}^{\phantom{kl}\ m}f_m^\epsilon=0\right)\vspace{1.5mm}&\quad, \\
\frac i2 \partial_{\beta_k}\, f_l^\epsilon+\tilde\varphi_k\,f_l^\epsilon-2\,\tilde \rho_{k\,l}^{\phantom{kl}\ m}f_m^\epsilon=0\label{l1pm}&\quad.
\end{array}\right .
\end{equation}
The general solution of this system now depends on two arbitrary constants $f^\epsilon_0$ and reads as : 
\begin{equation} \label{f1}
f_k^\epsilon=f^\epsilon\,\{x(y-z),\,y(z-x),\,z(x-y)\}
\end{equation}
with
\begin{equation}
f^\epsilon  =f_0^\epsilon\, Exp[-\frac 12(\frac 1x+\frac 1y+\frac 1z)](x\,y\,z)^{-\frac 34} \left((x-y)(y-z)(z-x)\right)^{-3/8}\qquad .
\end{equation} 
These solutions, as the previous ones, decay exponentially under the gravitational walls. Let us notice that, by contrast to what happens at level 0, here the solutions diverge on the symmetry walls, but not too abruptly : in a square integrable way. We mention in passing that previous works on the problem concluded to the absence of solutions at odd levels. \vspace{1.5mm}
{\par\noindent}
At the two next levels (2 and 3) the structure of the solutions becomes more involved (and more interesting). At \underline{level 2} we write the fifteen dimensional general solution as : 
\begin{equation}
\bm \Psi_{(2)}=1/2\sum_{\tiny \begin{array}{l}{\epsilon,\,\epsilon'=\pm}\\k,\,k'=1,2,3\end{array}}f^{\epsilon\,\epsilon'}_{k\,k'}\, \tilde b^k_\epsilon\tilde b^{k'}_{\epsilon'}\, \Psi_0\qquad \mbox{with}\qquad f^{\epsilon\,\epsilon'}_{k\,k'}=-f^{\epsilon'\,\epsilon}_{k'\,k}\qquad .
\end{equation}
Nine of these fifteen components can be expressed in closed form. They depend on three arbitrary constants. We obtain  for $\epsilon=\epsilon'=\pm$ :
\begin{equation}\label{fee}
 \{f_{1\,2}^{\epsilon\,\epsilon},f_{2\,3}^{\epsilon\,\epsilon},f_{3\,1}^{\epsilon\,\epsilon}\}=f^{\epsilon\,\epsilon}\{2(x\,y-y\,z-x\,z)+x\,y\,z,\ cyclic\ perm.\}
\end{equation}
with
\begin{eqnarray}
f^{\epsilon\,\epsilon}&=& Exp[-\frac 12(\frac 1x+\frac 1y+\frac 1z)](x\,y\,z)^{-3/4}(x-y)^{-1/8}(x-z)^{-1/8}(y-z)^{-1/8}\nonumber  \\
&& \left({ C_1}\,(x-z)^{-1/2}+{\epsilon\,C_2}\, (y-z)^{-1/2}\right)
\end{eqnarray}
When $\epsilon\neq\epsilon'$, the antisymmetric part of the amplitude of the wave function still leads to a discrete mode, depending on one constant : 
\begin{eqnarray}\label{fpmA}
&& \{f_{[1\,2]}^{+ -},f_{[2\,3]}^{+-},f_{[3\,1]}^{+-} \} =f^{+-}\{2(x\,y-y\,z-x\,z)+x\,y\,z,\ cyclic\ perm.\} \\
&&f^{+-}={C_3}\,Exp[-\frac 12(\frac 1x+\frac 1y+\frac 1z)]\,(x\,y\,z)^{-3/4}\left((y-z)(z-x)\right)^{-1/8}\,(x-y)^{-5/8}\nonumber\quad .
\end{eqnarray}
 However the six symmetric amplitudes $f^{+-}_{(k\,l)} =:k_{k\,l}$ are propagating modes!
They obey Maxwell-like equations :
\begin{eqnarray}\label{deltak}
\delta k\sim 0&:&\frac i2\partial^p k_{pa}+\phi^p\,k_{pa}-2\,\rho^{pq}_{\phantom{pq\,}}\,k_{pq}=0\\
d\,k\sim 0&:&\frac i2\partial_{[a}k_{b]c}-\phi_{[a}\,k_{b]c}+\mu_{ab}^{\phantom{ab}p}\,k_{pc}+2\,\rho_{[a\vert c\vert}^{\phantom{[a\vert c\vert}p}\,k_{b]p}=0\label{dk}
\end{eqnarray}
whose compatibility is guaranteed by Bianchi-like identities : $d^2=0=\delta^2$. 
They split into :\\
$\bullet$ five constraint equations (involving no ``time derivative'')\ ,\\
$\bullet$ six evolution equations (with respect to the $\beta$-time : $\beta^0=\beta^1+\beta^2+\beta^3$)\ .
{\par\noindent}
The general solution of Eqs [\ref{deltak}, \ref{dk}] is parametrized by two arbitrary functions of two variables (leaving in a plane $\beta^0=Cst$) from which we compute (via an Euler-Darboux-Poisson equation) the Cauchy data for the six $k_{ab}$, which then propagate thanks to the six evolution equations : $\partial_{\beta^0}\,k_{ab}=\dots $.\vspace{1.5mm}{\par\noindent}
A similar analysis can be done at \underline{level 3}. The twenty components of 
\begin{equation}\bm \Psi_{(3)}=\frac 1{\sqrt{2}}\sum_{\epsilon}
\frac 1{3!}f^{\epsilon}\eta_{p\,q\,r}\tilde b^p_{\epsilon} \tilde b^q_{\epsilon} \tilde b^r_{\epsilon}+\frac 1 {2}h^{\epsilon}_{p\,q ,r}\tilde b^p_{-\epsilon} \tilde b^q_{-\epsilon} \tilde b^r_{\epsilon}
\end{equation}
 split into two subsets (associated with the two values of $\epsilon$)  of ten each, which decouple in the constraint equations.  
Defining the dual components $h^{\epsilon}_{\phantom{\epsilon}a\,b}=\frac 12 \eta_a^{\phantom{a\,} p\,q}  h^{\epsilon}_{  p\,q  ,b}$,  we obtain that all modes may be expressed in terms of $h^{\epsilon}_{\phantom{\epsilon}(a\,b)}$ and a single function that parametrize $h^{\epsilon}_{\phantom{\epsilon}[a\,b]}$. These seven functions have to satisfy a (compatible) system of twelve first-order partial differential equations. The situation is similar to what we encountered at the previous level 2. Again we have equations of ``curl" and ``div" types. They lead to general solutions that for each value of the index $\epsilon$ depend on two arbitrary functions of two variables (plus one arbitrary constant). We refer the reader interested into the details to our works already quoted.\vspace{1.5mm}{\par\noindent}
On the solution space we have an up-down symmetry between solutions of level $N_F$ and those of level $6-N_F$. Having a solution at a given level $N_F$, by the transformation ($x\mapsto -x$, $y\mapsto -y$, $z\mapsto -z$) we obtain a solution leaving in the symmetric level $6-N_F$. This allows us to not discuss solutions at \underline{levels 4, 5 and 6}, except to mention that all the explicitly given solutions [\ref{f0}, \ref{f1}, \ref{fee}, \ref{fpmA}] are mapped on solutions that now grow exponentially into the gravitational walls. Accordingly 
we propose not to retain them.
\section{Quantum Evidences for a Hidden Kac-Moody Structure} 
The explicit expression of the Hamiltonian operator occurring in the anticommutation relations [\ref{SASB}] is :
\begin{equation}
\widehat  H =\frac12\left[ G^{ab} (\widehat \pi_a + i \, A_a)(\widehat \pi_b + i \, A_b) + \widehat \mu^2 + \widehat  W (\beta) \right]
\end{equation}
 where the vector potential term $A_a=\partial_a\ln[e^{\frac 34\beta^0}(\sinh\beta_{12}\,\sinh\beta_{23}\,\sinh\beta_{31})^{-\frac 18}]$ is pure gauge and thus can be eliminated by a multiplicative redefinition of the wave function $\bm \Psi$. The potential term $\widehat \mu^2 + \widehat  W (\beta)$ contains various interesting operators. Firstly let us decompose its $\beta$-depending part  into three parts according to :
\begin{equation}
\widehat  W (\beta) = \widehat W_g^{\rm bos.} (\beta) + \widehat  W_g^{\rm fer.} (\beta) + \widehat  W_{\rm sym}^{\rm fer.} (\beta) \qquad .
\end{equation}
The first one is spin independent, {\it i.e.} proportional to $\mbox{Id}_{64}$, and is (twice) the usual potential generating the chaotic mixmaster dynamic\cite{BKL,Misner:1969hg} of the Bianchi IX cosmological model :
\begin{equation}
\widehat W_g^{\rm bos.} (\beta)  = \frac12 \, e^{-4\beta^1} - e^{-2 (\beta^2 + \beta^3)} + {\rm cyclic}_{123}\qquad .
\end{equation}
The second one, that becomes singular on the symmetry walls, is  quartic in the fermionic operators. It involves the square of matrices representing a (reducible) $su(2)$ spin algebra :
\begin{equation}
\widehat  W_{\rm sym}^{\rm fer.} (\beta) = \frac12 \ \frac{(\widehat  S_{12} (\widehat \Phi))^2 - 1}{\sinh^2 (\beta_1-\beta_2)} + \mbox{ cyclic}_{123} \, ,
\end{equation}
 with
 \begin{equation}
\widehat  S_{12} (\widehat \Phi) =\frac12 [({\widehat {\overline\Phi}}\strut^3 \, \gamma^{\hat 0\hat 1\hat 2} (\widehat \Phi^1 + \widehat \Phi^2)) + ({\widehat {\overline\Phi}}\strut^1 \, \gamma^{\hat 0\hat 1\hat 2} \, \widehat \Phi^1)  
+({\widehat {\overline\Phi}}\strut^2 \, \gamma^{\hat 0\hat 1\hat 2} \, \widehat \Phi^2) - ({\widehat {\overline\Phi}}\strut^1\, \gamma^{\hat 0\hat 1\hat 2} \, \widehat \Phi^2)] \, .
\end{equation}
The last one  involves the gravitational walls
 \begin{equation}
\widehat  W_g^{\rm fer.} (\beta , \widehat \Phi) =e^{-\alpha_{11}^g (\beta)}  \ \widehat{\!\!J}_{11} (\widehat \Phi) + e^{-\alpha_{22}^g (\beta)} \ \widehat{\!\!J}_{22} (\widehat \Phi) +e^{-\alpha_{33}^g (\beta)} \ \widehat{\!\!J}_{33} (\widehat \Phi) \, . \nonumber
\end{equation}
where
\begin{equation}
\widehat{\!\!J}_{11} (\widehat \Phi) = \frac12 \, [{\widehat {\overline\Phi}}\strut^1\gamma^{\hat 1\hat 2\hat 3} (4 \widehat \Phi^1 + \widehat \Phi^2 + \widehat \Phi^3) +{\widehat {\overline\Phi}}\strut^2 \, \gamma^{\hat 1\hat 2\hat 3} \, \widehat \Phi^3] \, .
\end{equation}
Remarkably the set 
$\widehat J_{\alpha_i}=\{\widehat  S_{12}$, $\widehat  S_{23}$, and $\widehat{\!J}_{11}\}$ generates (via commutators) a (reducible) $64$-dimensional representation of the (infinite-dimensional) ``maximally compact'' sub-algebra $K(AE_3) \subset AE_3$.  The following
exponentials of these operators, $\widehat{\mathcal R}_{\alpha_i}:=Exp[\pm i \pi/2\,\widehat J_{\alpha_i}]$, that verify $(\widehat{\mathcal R}_{\alpha_i})^8=\mbox{Id}$,  generate a spinorial extension\cite{Damour:2009zc} of $AE_3$.
{\par\noindent}
Far from the wall the potential reduces to $\widehat\mu^2$. Classically this term is zero. Quantum mechanically it is an operator that is quartic in the fermionic fields and given by 
\begin{equation}
\widehat\mu^2=\frac 12-\frac 78 \,\widehat C_F^2=\frac 12-\frac 78 \,\left ( \frac 12 G_{ab}\overline{\widehat \Phi^a}\gamma^{123}\widehat\Phi^b\right)^2 \qquad .
\end{equation}
 This term plays the r\^ole of a mass for the particule representing the evolution of the wave function of the Universe in a short wave (WKB) limit. Its  eigenvalues are : $(-59/8\vert_2,\,-3\vert_{12},\,-3/2\vert_{30},\,+1/2\vert_{20})$, and lead to an asymptotically spacelike motion (in $\beta$ space) for more than a half of the continuous solution subspaces.

\section{Conclusions}
We have obtain a description of the space of solution of the constraint equations resulting from the quantization of the $\mathcal N=1$ $d=4$ simple supergravity, in the framework of the minisuperspace defined by the Bianchi IX cosmological model. This space appears as filled both with discrete modes {\it i.e.} solutions depending only on a finite number of constants, and continuous modes {\it i.e.} solutions depending on arbitrary functions. These solutions are labeled by a fermionic number $N_F$ that ranges from 0 to 6. At level zero we have a unique ground state, describing a localised Planckian size Universe, whose wave function decays in the gravitational walls and vanishes on the symmetry walls. At level $N_F=1$ the wave function is similar, excepted that it blows up on the symmetry walls, but not too drastically, remaining square integrable. At level $N_F=2$ we obtain again such solutions but also structurally different ones, depending on arbitrary functions that define initial data. At level $N_F=3$ also the solutions depend on arbitrary functions (and one constant). The discrete solutions at level higher that $N_F=3$ appears not to be physically acceptable : they grow behind the gravitational walls, a classically forbidden region. In the limit of small wavelength, the wave function bounces on the various walls. The associated reflection operators define a spinorial extension of the Weyl group of the $AE_3$ Kac-Moody algebra. Finally, far from the wall,  the Wheeler DeWitt equation involves, contrary to the classical picture described by a {\it massless} bouncing particle, a squared-mass term that, for 44 among the 64 components of the wave function is negative, corresponding to a tachyonic motion that permits to the wave-packet to bounce back in $\beta^0$-time and so to avoid the cosmological singularity located at $\beta^0=\infty$. This results from a specific quantum effect due to the term quartic in the fermion that we have rigorously taken into account during  all our analysis.
 
\section*{Acknowledgments} Ph. S. thanks IHES for its kind hospitality; his work has been partially supported by ``Communaut\'e fran\c caise de Belgique -- Actions de Recherche concert\'ees'' and by IISN-Belgium (convention 4.4511.06).
\bibliography{MyBiblio}{}

\end{document}